\documentclass[journal]{IEEEtran}

\ifCLASSINFOpdf
\else
\usepackage[dvips]{graphicx}
\fi
\usepackage{url}
\usepackage{graphicx}

\usepackage{color}
\usepackage{balance}
\usepackage{bm}
\usepackage{amsmath}
\usepackage{amssymb}
\usepackage{lipsum}
\usepackage{algorithm}
\usepackage{algorithmic}
\usepackage{multirow}
\usepackage{booktabs}
\usepackage{array}
\usepackage{lipsum}
\usepackage{enumerate}
\usepackage{stfloats}
\usepackage{subfigure}
\usepackage{cases}
\usepackage{diagbox}
\usepackage{cite}
\usepackage{enumitem}
\usepackage{balance}
\usepackage{hyperref} 
\usepackage{amsthm}

\theoremstyle{plain}
\newtheorem{defi}{Definition}

\newtheorem{theorem}{Theorem}
\newtheorem{lemma}{Lemma}

\newtheorem{insight}{Insight}

\allowdisplaybreaks[4]

\begin{document}
	
	\title{Edge AI Inference in ISCC Networks: Sensing Accuracy Analysis and Precoding Design}
	
	\author{Lingyun Xu, \IEEEmembership{Graduate Student Member,~IEEE}, Bowen Wang, \IEEEmembership{Graduate Student Member,~IEEE},\\ Huiyong Li and Ziyang Cheng,~\IEEEmembership{Senior Member,~IEEE}
		\vspace{-1em}
		\thanks{L. Xu, H. Li, and Z. Cheng are with the School of Information and Communication Engineering, University of Electronic Science and Technology of China, 611731, Chengdu, China. (email: xusherly@std.uestc.edu.cn, \{hyli, zycheng\}@uestc.edu.cn).
			B. Wang is with the Department of Engineering, King's College London, London, WC2R 2LS, UK. (email: bowen.wang@kcl.ac.uk).}
	}
	
	\maketitle
	
	\begin{abstract}
		This work explores the relationship between sensing accuracy and precoding coefficients for edge artificial intelligence (AI) inference in integrated sensing, communication and computation (ISCC) networks.
		We start by constructing a system model of an over-the-air-empowered ISCC network for edge AI inference, involving distributed edge sensors for feature extraction and an edge server for classification.
		Based on this model, we introduce a discriminant gain (DG) to characterize sensing accuracy and novelly derive an explicit function of the DG about precoding coefficients, giving valuable insights into precoding design.
		Guided by this, we propose an effective precoding algorithm to solve a non-convex DG-maximization problem.
		Simulation results demonstrate that the proposed design achieves up to $15\%$ and $10\%$ sensing accuracy improvements on synthetic and real-world datasets, respectively, over the conventional scheme at low SNR, thereby validating its effectiveness and superiority for edge AI inference in ISCC networks.
		
	\end{abstract}

	
	\IEEEpeerreviewmaketitle
	
	\vspace{-1.2em}
	\section{Introduction}
	\vspace{-0.2em}
	
	\IEEEPARstart{T}{he} forthcoming sixth-generation (6G) networks aim to support ubiquitous edge-born \textit{Artificial Intelligence} (AI) \cite{li2019edge,8970161,letaief2021edge}, driving innovations in autonomous systems and the metaverse \cite{10445209,zhu2023pushing,10024766,wen2024survey}. 
	Supporting these services necessitates a synergistic integration of sensing, communication, and computation, leading to the development of the \textit{Integrated Sensing, Communication, and Computation} (ISCC) framework \cite{zhu2023pushing,10024766,wen2024survey}. 
	Within this context, \textit{edge AI} inference is a key enabler \cite{wang2020convergence}, as it deploys pre-trained AI models on edge devices or servers to process real-time sensing data, thereby reducing latency and improving resource utilization and data privacy \cite{li2019edge,8970161,letaief2021edge,10445209,zhu2023pushing,10024766,wen2024survey,wang2020convergence,Meng2024WCM,Zhang2025WCM}.

	Extensive research has been dedicated to optimizing edge AI inference.
	For instance, authors in \cite{lee2023wireless} proposed an adaptive split inference method for edge intelligence, where split points of deep neural networks were selected dynamically.
	Another work \cite{yao2025energy} extended \cite{lee2023wireless} by considering the data collection process.
	Moreover, \cite{zhang2022accelerating} proposed a joint design of beamforming and time allocation to accelerate edge intelligence.
	Despite these advancements, these works \cite{lee2023wireless,yao2025energy,zhang2022accelerating} primarily focus on scenarios involving a single edge sensor, which may struggle with limited sensing coverage, narrow sensing view, and high vulnerability to blockage or interference.
	
	
	To address these limitations, leveraging distributed sensors has emerged as a promising paradigm to enhance sensing accuracy and robustness, since multi-view observations can effectively overcome limited coverage and signal blockage \cite{wang2025ultra,zhuang2023integrated,yang2025mimo}.
		For example, \cite{wang2025ultra} developed an ultra-low-latency inference framework for distributed sensors with packet length-aware sensing accuracy analysis.
		In multi-sensor edge AI, the server needs to aggregate distributed features for inference, which is a summation operation.
		To this end, \textit{Over-the-air Computation} (AirComp) enables simultaneous transmission and aggregation by exploiting the superposition property of wireless channels.
		Recent works have incorporated AirComp into ISCC systems.
		For example, \cite{zhuang2023integrated} proposed a task-oriented ISCC scheme with AirComp for a fair classification task, and \cite{yang2025mimo} further improved transmission efficiency by utilizing multiple-input multiple-output AirComp.

	Among these works \cite{wang2025ultra,zhuang2023integrated,yang2025mimo}, \textit{Discriminant Gain} (DG) is utilized to quantify class discernibility. 
	However, for analytical tractability, existing studies \cite{wang2025ultra,zhuang2023integrated,yang2025mimo} analyzed DG based on conventional communication-centric \textit{Zero-Forcing} (ZF) and \textit{Maximum-Ratio Transmission} (MRT). 
	Intuitively, the precoder is pivotal in shaping the feature space. 
	By strictly adhering to the ZF and MRT principles, the \textit{degrees of freedom} (DoF) for signal enhancement are constrained, potentially degrading the DG and inference accuracy.
	Driven by this, it is imperative to re-examine the coupling between precoding coefficients and inference performance. 
	In this letter, we consider a multi-sensor AirComp-empowered ISCC system for edge AI inference. The focus of this work is on the following questions:
	\textit{Can we characterize the explicit relationship between precoding coefficients and DG in multi-sensor ISCC? To what extent can a tailored precoder design enhance edge AI inference performance compared to conventional approaches?}
	
	We provide comprehensive answers to both questions, with our main contributions summarized as follows:
	\textit{First}, we establish an explicit analytical relationship between the precoding coefficients and the DG. 
	This clarifies the impact of signal processing on sensing performance and enables a more concise DG-oriented optimization framework.
	\textit{Second}, leveraging this analysis, we formulate a DG-maximization problem to design the precoder.
	This facilitates an efficient solution without resorting to communication-centric heuristic simplifications, thereby ensuring superior sensing and inference accuracy.
	\textit{Third}, simulation results on both synthetic and real-world datasets demonstrate the effectiveness of our design, confirming that tailored precoding significantly boosts the inference accuracy in the ISCC networks.
	
	\vspace{-0.8em}
	\section{System Model}
	\vspace{-0.2em}
	
	\begin{figure}[!t]
		\centering 
		\includegraphics[width=1\linewidth]{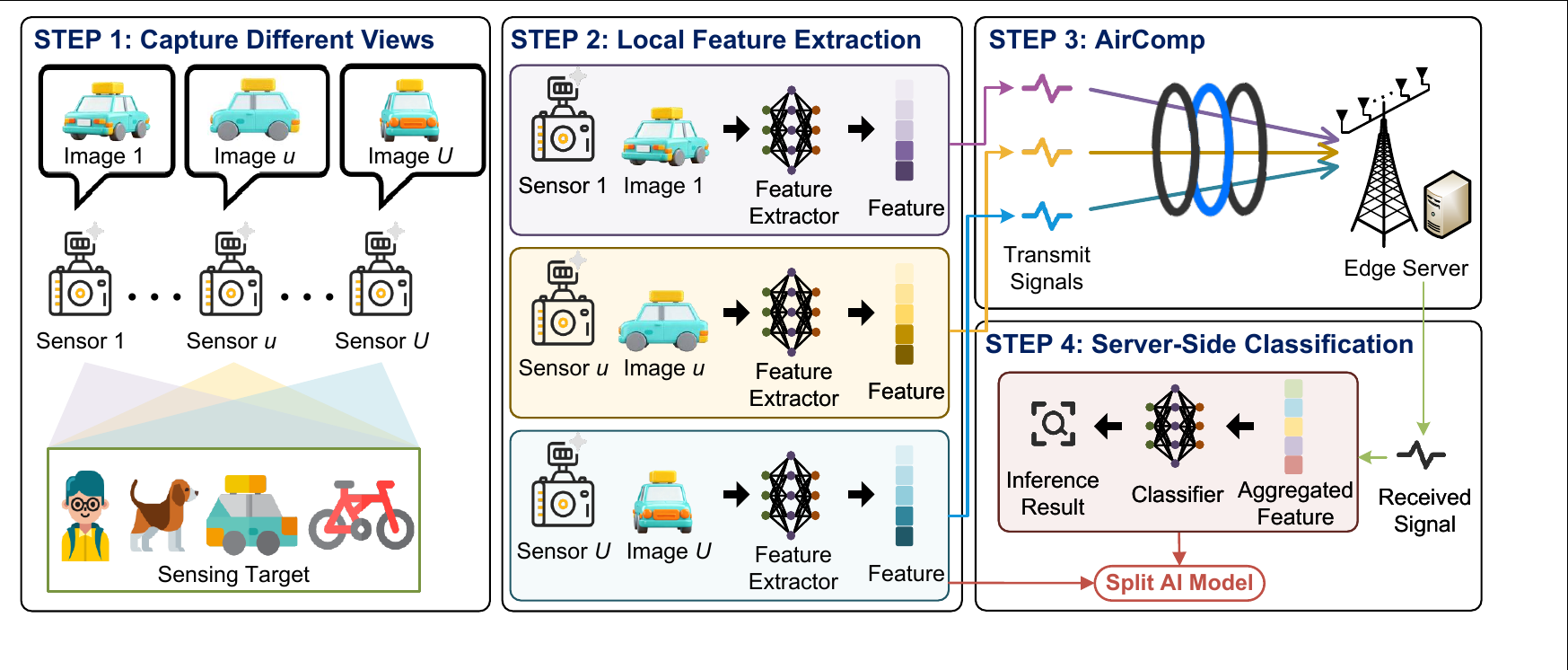}
		\vspace{-2em}
		\caption{An overview of the AirComp-empowered multi-sensor edge AI inference framework.
			The pipeline consists of four integrated stages: 
			(1) Multi-View Sensing: $U$ distributed sensors capture diverse views of the target; 
			(2) Local Feature Extraction: Sensor-side AI models extract low-dimensional features from raw data; 
			(3) Over-the-Air Computation: Local features are transmitted and wirelessly aggregated through AirComp; 
			(4) Server-side Classification: The edge server receives the aggregated signal and the server-side AI model performs final inference.
		}
		\label{iscc_model}
		\vspace{-1.5em}
	\end{figure}

	Fig. \ref{iscc_model} illustrates the architecture of the considered ISCC network that facilitates edge AI inference for classification tasks.
	We consider a classification task with $L$ classes. 
	The network consists of $U$ distributed single-antenna sensors.
	These sensors capture distinct views of a sensing target, which belongs to a common object class $l$, and employ a sensor-side AI model for local feature extraction.
	Subsequently, local features at $U$ distributed sensors are transmitted uplink over $K$ subcarriers to the edge server, which is an $M$-antenna access point that aggregates the features via AirComp.
	The final classification is then carried out by a server-side AI model based on the aggregated features, generating the ultimate inference result.
	
	\vspace{-0.8em}
	\subsection{Data Distribution Models}
	\vspace{-0.2em}
	
	At the $u$-th sensor, the local feature vector ${{\bf{x}}_u} = {\left[ {{x_{u,1}}, \ldots ,{x_{u,N}}} \right]^T} \in {{\mathbb R}^N}$ is generated by a pre-trained sensor-side AI model from the captured image of the sensing target.
	Following \cite{lan2022progressive,zhuang2023integrated,yang2025mimo}, the feature vectors are statistically modeled by a Gaussian mixture distribution \cite{mclachlan2014number}.
	Thus, the joint \textit{Probability Density Function} (PDF) of all local feature vectors, $\left( {{{\bf{x}}_1}, \ldots ,{{\bf{x}}_U}} \right)$, is expressed as
	\begin{equation}\label{eq:1}
		f\left( {{{\bf{x}}_1}, \ldots ,{{\bf{x}}_U}} \right) =  \frac{1}{L} \textstyle \sum\limits_{l = 1}^L {\prod\limits_{u = 1}^U {f\left( {{{\bf{x}}_u}|l} \right)} },
	\end{equation}
	where $f\left( {{{\bf{x}}_u}|l} \right) = {\cal N}\left( {{{\bm{\mu }}_l},{\bm{\Sigma }}} \right)$ denotes the Gaussian PDF for the class $l$, with the centroid ${{\bm{\mu }}_l} = {\left[ {{\mu _{l,1}}, \ldots ,{\mu _{l,N}}} \right]^T} \in {{\mathbb R}^N}$ and the covariance ${\bm{\Sigma }} = {\rm{diag}}( {\sigma _1^2, \ldots ,\sigma _N^2} ) \in {{\mathbb R}^{N \times N}}$.
	Note that the centroid and covariance are determined a priori during the training stage using sufficient sample data \cite{zhuang2023integrated,liu2024energy,wang2025ultra,yang2025mimo}.

	\vspace{-0.8em}
	\subsection{Over-the-Air Computation Model}
	\vspace{-0.2em}
	
	Following local feature extraction by the sensor-side AI model, the aggregation of these feature vectors at the edge server is performed using the AirComp technique. 
	We detail the implementation through the following transmission and reception operations.
	
	On the local sensor side, to enable the transmission of the $N$-dimensional real-valued feature vector $\mathbf{x}_u$ over $K$ subcarriers, we first map the real-valued elements to a complex-valued symbol. 
	Specifically, with $N = 2K$, the transmit symbol from the $u$-th sensor on the $k$-th subcarrier is defined as
	\begin{equation}\label{eq:2}
		{s_{u,k}} = {x_{u,2k - 1}} + \jmath {x_{u,2k}}.
	\end{equation}
	Accordingly, the signal transmitted by the $u$-th sensor at the $k$-th subcarrier is given by
	\begin{equation}\label{eq:3}
		{{{x}}_{{\rm{TX}},u,k}} = {v_{u,k}}{s_{u,k}},
	\end{equation}
	where ${v_{u,k}}$ denotes the corresponding precoding coefficient.
	
	On the edge server side, the received signal at the $k$-th subcarrier is processed by a linear receive precoder ${{\bf{w}}_k} \in {{\mathbb C}^M}$. 
	The estimated complex symbol corresponding to the aggregated feature component at the $k$-th subcarrier can be thus expressed as
	\begin{equation}\label{eq:4}
		{\hat s_k} = {\bf{w}}_k^H \textstyle\sum\limits_{u = 1}^U {{{\bf{h}}_{u,k}}{v_{u,k}}{s_{u,k}}}  + {\bf{w}}_k^H{{\bf{n}}_k},
	\end{equation}
	where ${{\bf{h}}_{u,k}} \in {{\mathbb C}^M}$ denotes the communication channel from the $u$-th sensor to the edge server at the $k$-th subcarrier \cite{liu2014channel}, and ${{\bf{n}}_k} \sim {\cal C}{\cal N}\left( {{\bf{0}},\sigma _{\rm{c}}^2{\bf{I}}} \right)$ denotes additive white Gaussian noise.
	
	Finally, by demultiplexing the real and imaginary parts of the estimated symbols based on the modulation scheme in \eqref{eq:2}, the aggregated feature vector ${\bf{\hat x}} = \left[ {{{\hat x}_1}, \ldots ,{{\hat x}_N}} \right]^T \in {{\mathbb R}^N}$ is constructed with
	\begin{equation}\label{eq:5}
		{{\hat x}_{2k - 1}} = \Re \left\{ {{{\hat s}_k}} \right\}, {{\hat x}_{2k}} = \Im \left\{ {{{\hat s}_k}} \right\},\forall k.
	\end{equation}
	
	\vspace{-1.3em}
	\section{Sensing Accuracy Analysis}\label{sec:3}
	\vspace{-0.2em}
	
	Having illustrated the workflow of the ISCC network, the next crucial step is to evaluate the system sensing performance. 
	To this end, we utilize the DG metric to rigorously quantify the sensing accuracy. 
	This section establishes the analytical relationship between the DG and precoding coefficients.
	
	\vspace{-1.1em}
	\subsection{Discriminant Gain}\label{sec:DG}
	\vspace{-0.2em}

	Since the overall sensing accuracy is fundamentally determined by the discernibility between any two classes $(l, l')$, we adopt the DG as the key performance metric, formally defined below.
	\begin{defi}\label{def:1}
		Assuming that ${\hat x_n}$ and ${\hat x_m}$ are independent for any $n \ne m$, the DG of the class pair $\left(l,l'\right)$ with $l \ne l'$ is defined by\cite{lan2022progressive}
		\begin{equation}\label{eq:6}
			\begin{aligned}
				&  \!{G_{l,l'}}\left( {{\bf{\hat x}}} \right) = \textstyle \sum\limits_{n = 1}^N {{G_{l,l'}}\left( {{{\hat x}_n}} \right)}  \\
				& = \textstyle\sum\limits_{n = 1}^N {\rm{KL}}( {f( {{{\hat x}_n}|l} )\left\| {f\left( {{{\hat x}_n}|l'} \right)} \right.} ) + \sum\limits_{n = 1}^N {\rm{KL}}( {f( {{{\hat x}_n}|l'} )\left\| {f( {{{\hat x}_n}|l} )} \right.} ) ,
			\end{aligned}
		\end{equation}
		where ${\rm{KL}} ( \pi_1 || \pi_{2})$ denotes the Kullback-Leibler divergence between two distributions $\pi_1$ and $\pi_{2}$, and $f\left( {{{\hat x}_n}|l} \right)$ represents the PDF of the $n$-th-dimension feature ${{{\hat x}_n}}$ of the $l$-th class.
	\end{defi}
	
	Based on \textbf{Definition \ref{def:1}}, the overall DG is calculated by averaging all the pair-wise DGs
	\begin{equation}\label{eq:7}
		G\left( {{\bf{\hat x}}} \right) = \frac{2}{{L\left( {L - 1} \right)}} \textstyle\sum\limits_{l' = 1}^L {\sum\limits_{l < l'} {{G_{l,l'}}\left( {{\bf{\hat x}}} \right)} } .  
	\end{equation}
	Note that a higher DG leads to a higher inference accuracy, namely, a higher sensing accuracy.
	
	\vspace{-1.1em}
	\subsection{Aggregated Feature Statistical Properties}
	\vspace{-0.2em}
	
	The DG $G(\hat{\mathbf{x}})$ in \eqref{eq:7} fundamentally depends on the distribution of the aggregated features. 
	To characterize the impact of the precoding coefficients on the DG, it is essential to analyze the statistical properties of the aggregated feature vector $\hat{\mathbf{x}}$. 
	The distribution of $\hat{\mathbf{x}}$, which results from the AirComp aggregation of $U$ local feature vectors at the edge server, is provided by the following lemma.
	\begin{lemma}\label{lemma:1}
		Under the local feature distribution \eqref{eq:1}, the aggregated feature \eqref{eq:5}, and the assumption $\sigma_{2k-1}^2 = \sigma_{2k}^2 = \delta_k^2, \;\forall k$, the PDF of the aggregated feature ${{\bf{\hat x}}}$ of the $l$-th class can be expressed as
		\begin{equation}\label{eq:8}
			f\left( {{\bf{\hat x}}|l} \right) = {\cal N}( {{{{\bm{\hat \mu }}}_l},{\bf{\hat \Sigma }}} ),
		\end{equation}
		where the centroid ${{{\bm{\hat \mu }}}_l} = {\left[ {{{\hat \mu }_{l,1}}, \ldots ,{{\hat \mu }_{l,N}}} \right]^T} \in {{\mathbb R}^N}$ is given by
		\begin{equation}
			\begin{aligned}
				{{\hat \mu }_{l,2k - 1}} =&  \textstyle\sum\limits_{u = 1}^U {\left( {{a_{u,k}}{\mu _{l,2k - 1}} - {b_{u,k}}{\mu _{l,2k}}} \right)},\forall k, \\
				{{\hat \mu }_{l,2k}} =&  \textstyle\sum\limits_{u = 1}^U {\left( {{a_{u,k}}{\mu _{l,2k}} + {b_{u,k}}{\mu _{l,2k - 1}}} \right)} ,\forall k,
			\end{aligned}
		\end{equation}
		and the covariance ${\bf{\hat \Sigma }} = {\rm{diag}}( {{\hat \Sigma }_{1,1}},\dots,{{\hat \Sigma }_{N,N}} ) \in {{\mathbb R}^{N \times N}}$ is
		\begin{equation}
			\begin{aligned}
				{{\hat \Sigma }_{2k - 1,2k - 1}} & = {{\hat \Sigma }_{2k,2k}} \\
				& = \delta _k^2 {\textstyle\sum\limits_{u = 1}^U {{{\left| {{\omega _{u,k}}} \right|}^2}}}  + \frac{{\sigma _{\rm{c}}^2}}{2}\left\| {{{\bf{w}}_k}} \right\|_F^2 \buildrel \Delta \over = {\eta _k},\forall k,
			\end{aligned}
		\end{equation}
		with ${\omega _{u,k}} = {\bf{w}}_k^H{{\bf{h}}_{u,k}}{v_{u,k}}, {a_{u,k}} = \Re \left\{ {{\omega _{u,k}}} \right\},{b_{u,k}} = \Im \left\{ {{\omega _{u,k}}} \right\}$.
	\end{lemma}
	\begin{IEEEproof}
		Please refer to the supplementary material (SM) Appendix \hyperref[app:A]{A}. 
	\end{IEEEproof}
	
	\subsection{Effect of Precoding on Discriminant Gain}
	By leveraging the definition of the DG (\textbf{Definition \ref{def:1}}) and the statistical properties of the aggregated feature distribution (\textbf{Lemma \ref{lemma:1}}), the analytical relationship between the DG and precoding coefficients is established in the following theorem.
	\begin{theorem}\label{th:1}
		The pair-wise DG of the class pair $\left(l,l'\right)$ with $l \ne l'$ can be expressed as an explicit function of the precoding coefficients, $\{ {v_{u,k}}\} ,\{ {{\bf{w}}_k}\} $, as follows.
		\begin{equation}\label{eq:9}
			{G_{l,l'}}\left( \{ {v_{u,k}}\} ,\{ {{\bf{w}}_k}\} \right)  = \textstyle\sum\limits_{k = 1}^K { {\frac{{{{\left| {\sum\limits_{u = 1}^U {{\bf{w}}_k^H{{\bf{h}}_{u,k}}{v_{u,k}}} } \right|}^2}{\psi _{l,l',k}}}}{{\delta _k^2\sum\limits_{u = 1}^U  {{{\left| {{\bf{w}}_k^H{{\bf{h}}_{u,k}}{v_{u,k}}} \right|}^2}}  +  \frac{\sigma _{\rm{c}}^2}{2}\left\| {{{\bf{w}}_k}} \right\|_F^2}}}},
		\end{equation}
		where ${\psi _{l,l',k}} = {\left( {{\mu _{l,2k - 1}} - {\mu _{l',2k - 1}}} \right)^2} + {\left( {{\mu _{l,2k}} - {\mu _{l',2k}}} \right)^2}$.
	\end{theorem}
	\begin{IEEEproof}
		Based on \textbf{Definition \ref{def:1}} and \textbf{Lemma \ref{lemma:1}}, the pair-wise DG can be further rewritten as
		\begin{equation}\label{eq:10}
			\begin{aligned}
				{G_{l,l'}}\left( {{\bf{\hat x}}} \right)  = &  \textstyle\sum\limits_{k = 1}^K {\eta _k} {{\left( {{{\hat \mu }_{l,2k - 1}} - {{\hat \mu }_{l',2k - 1}}} \right)}^2} \\
				& + \textstyle\sum\limits_{k = 1}^K {\eta _k} {{\left( {{{\hat \mu }_{l,2k}} - {{\hat \mu }_{l',2k}}} \right)}^2} .
			\end{aligned}
		\end{equation}
		By performing equivalent transformations, we further have
		\begin{subequations}
			\begin{align}
				{{\hat \mu }_{l,2k - 1}} - {{\hat \mu }_{l',2k - 1}} &=  \textstyle\sum\limits_{u = 1}^U {\left( {{a_{u,k}}\Delta {\mu _{l,l',2k - 1}} - {b_{u,k}}\Delta {\mu _{l,l',2k}}} \right)} \nonumber\\
				&=\Re \left\{ { \textstyle\sum\limits_{u = 1}^U {{\omega _{u,k}}\Delta {d_{l,l',k}}} }\right\},\label{eq:11a}\\
				{{\hat \mu }_{l,2k}} - {{\hat \mu }_{l',2k}} &=  \textstyle\sum\limits_{u = 1}^U {\left( {{a_{u,k}}\Delta {\mu _{l,l',2k}} - {b_{u,k}}\Delta {\mu _{l,l',2k - 1}}} \right)} \nonumber\\
				&= \Im \left\{ { \textstyle\sum\limits_{u = 1}^U {{\omega _{u,k}}\Delta {d_{l,l',k}}} } \right\},\label{eq:11b}
			\end{align}
		\end{subequations}
		where we define $\Delta {\mu _{l,l',2k - 1}} = {\mu _{l,2k - 1}} - {\mu _{l',2k - 1}},\Delta {\mu _{l,l',2k}} = {\mu _{l,2k}} - {\mu _{l',2k}}$ and $\Delta {d_{l,l',k}} = {\Delta {\mu _{l,l',2k - 1}} + \jmath \Delta {\mu _{l,l',2k}}}$.
		
		Combing \eqref{eq:11a} and \eqref{eq:11b}, we can derive
		\begin{equation}\label{eq:12}
			\begin{aligned}
				{G_{l,l'}}\left( {{\bf{\hat x}}} \right)  =& \textstyle \sum\limits_{k = 1}^K { {{\eta _k} {\left| { \textstyle\sum\limits_{u = 1}^U {{\omega _{u,k}}\Delta {d_{l,l',k}}} } \right|^2} } }\\
				=& \textstyle\sum\limits_{k = 1}^K { {\frac{{{{\left| {\sum\limits_{u = 1}^U {{\bf{w}}_k^H{{\bf{h}}_{u,k}}{v_{u,k}}} } \right|}^2}{\left| {\Delta {d_{l,l',k}}} \right|^2}}}{{\delta _k^2\sum\limits_{u = 1}^U  {{{\left| {{\bf{w}}_k^H{{\bf{h}}_{u,k}}{v_{u,k}}} \right|}^2}}  +  \frac{\sigma _{\rm{c}}^2}{2}\left\| {{{\bf{w}}_k}} \right\|_F^2}}}} .
			\end{aligned}
		\end{equation}
		
		The proof is completed.
	\end{IEEEproof}
	From \textbf{Theorem \ref{th:1}}, we can obtain the following insight.
	\begin{insight}{(Design Guideline)}\label{in:1}
		A larger DG is desirable as it directly correlates with higher inference accuracy, thereby enhancing sensing performance. 
		Eq. \eqref{eq:9} facilitates DG-oriented optimization via tailored precoding design. 
		Consequently, our analysis bridges the gap in current literature by providing an interpretable and theoretically grounded perspective on how precoding shapes discriminability between different classes.
	\end{insight}
	
	\vspace{-1.5em}
	\section{Proposed Joint Precoding Design}\label{sec:4}
	\vspace{-0.2em}
	
	Motivated by \textbf{Insight \ref{in:1}}, to improve sensing accuracy, we formulate the precoding design as a DG-maximization problem as follows
	\begin{subequations}
		\begin{align}
			&\mathop {\max }\limits_{\{ {v_{u,k}}\} ,\{ {{\bf{w}}_k}\} } \;\frac{2}{{L\left( {L - 1} \right)}}\textstyle\sum\limits_{l' = 1}^L {\sum\limits_{l < l'} {{G_{l,l'}}\left( {\{ {{\bf{w}}_k}\} ,\{ {v_{u,k}}\} } \right)} } \label{eq:P1-a}\\
			&\qquad {\rm{s.t.}}\quad\;\; \textstyle\sum\limits_{k = 1}^K {{{\left| {{v_{u,k}}} \right|}^2}}  \le \tilde P_u^{{\rm{sensor}}},\forall u \label{eq:P1-b}\\
			&\qquad \qquad \quad\left\| {{{\bf{w}}_k}} \right\|_F^2 \le P_k^{{\rm{server}}},\forall k,\label{eq:P1-c}
		\end{align}\label{eq:P1}%
	\end{subequations}
	where \eqref{eq:P1-a} defines the objective function, which is the overall DG to be maximized for the enhancement of sensing accuracy.
	Constraint \eqref{eq:P1-b} imposes a limit on the total transmit power of the $u$-th sensor excluding contribution of transmit symbol variance, ensuring it remains below the threshold $\tilde P_u^{\mathrm{sensor}}$.
	Constraint \eqref{eq:P1-c} specifies the receive power constraint, $P_k^{{\rm{server}}}$, at the edge server corresponding to the $k$-th subcarrier.
	
	Note that the problem \eqref{eq:P1} is non-convex due to the fractional objective function.
	To solve this problem effectively, we propose an alternating optimization-based algorithm, which proceeds according to the following two steps
	\begin{align}
		\mathop {\max }\limits_{{{\bf{w}}_k}} & \; \frac{{{{| {{\bf{w}}_k^H{{\bf{g}}_k}} |}^2}}}{{\delta _k^2{\bf{w}}_k^H{{\bf{R}}_k}{{\bf{w}}_k} + \frac{1}{2}\sigma _{\rm{c}}^2\| {{{\bf{w}}_k}} \|_F^2}}, \;\; {\rm{s.t.}} \; \eqref{eq:P1-c}, \label{eq:n16} \\
		\mathop {\max }\limits_{\{ {v_{u,k}}\} } & \; \textstyle\sum\limits_{l' = 1}^L {\sum\limits_{l < l'} {\sum\limits_{k = 1}^K {\left( {\frac{{{{\left| {\sum\limits_{u = 1}^U {{\alpha _{u,k}}{v_{u,k}}} } \right|}^2}{\psi _{l,l',k}}}}{{\sum\limits_{u = 1}^U {{\beta _{u,k}}{{\left| {{v_{u,k}}} \right|}^2}}  + {\gamma _k}}}} \right)} } } , \;\; {\rm s.t.} \; \eqref{eq:P1-b}, \label{eq:n17} 
	\end{align}
	The definition of variables $\{\mathbf{g}_k\}$, $\{\mathbf{R}_k\}$, $\{ \alpha_k \}$, and $\{ \beta_k \}$, solutions to the problem \eqref{eq:P1}, convergence analysis and complexity comparison are given in SM Appendix.

	\vspace{-1em}
	\section{Simulation Results}
	\vspace{-0.2em}
	
	In this section, simulation results are provided to validate the effectiveness and feasibility of the proposed design scheme.
	
	\vspace{-1.3em}
	\subsection{System Setup}
	\vspace{-0.2em}
	
	\subsubsection{Parameter Settings}
	All simulation parameters and dataset configurations follow the settings established in prior works \cite{wang2025ultra,zhuang2023integrated,yang2025mimo}.
	We assume that there exist $U=12$ distributed sensors and an edge server equipped with $M=10$ antennas.
	The number of subcarriers is set as $K=10$ and the dimension of the local feature vector is set as $N=2K=20$.
	Besides, the receive power is assumed as $P_k^{{\rm{server}}} = 30{\rm{dBm}},\forall k$, and the communication noise power is assumed as $\sigma _{\rm{c}}^2 = 30{\rm{dBm}}$.
	To facilitate the simulation setup, we introduce an analytical equivalent SNR, ${{\rm{SNR}}} = \hat \varpi U\tilde P_u^{{\rm{sensor}}}/K\sigma _{\rm{c}}^2$, to flexibly determine the transmit power excluding the transmit symbol variance, $\tilde P_u^{{\rm{sensor}}},\forall u$, for each sensor.
	Specifically, we define $\hat \varpi  = \mathop {\max }\nolimits_k {\varpi _k}$, where $ {\varpi _k} = \sigma _{2k - 1}^2 + \sigma _{2k}^2 + \frac{1}{L}\sum\nolimits_{l = 1}^L {( {\mu _{l,2k - 1}^2 + \mu _{l,2k}^2} )},\forall k$.
	Moreover, we let $\delta _k^2 = (\sigma _{2k - 1}^2 + \sigma _{2k}^2)/2,\forall k$.
	
	\subsubsection{Inference Tasks and Models}
	
	We consider two split inference tasks and corresponding models as follows.
	\begin{itemize}
		\item \textbf{Linear Classification on Synthetic Gaussian Mixture Data:} The synthetic Gaussian mixture data are generated by setting the centroid, ${{\bm{\mu }}_l},\forall l$, from the $(N(l-1)/L)$-th dimension to the $(Nl/L)$-th dimension as $-1$ and other elements as $+1$, and the covariance matrix as ${\bf{\Sigma }} = 3{{\bf{I}}_N}$ \cite{wang2025ultra}.
		The linear classifier is based on the Mahalanobis distance:
		\begin{equation}
			\hat l = \mathop {\arg \min }\nolimits_l \sqrt {{{\left( {{\bf{\hat x}} - {{{\bm{\hat \mu }}}_l}} \right)}^T}{{{\bf{\hat \Sigma }}}^{ - 1}}\left( {{\bf{\hat x}} - {{{\bm{\hat \mu }}}_l}} \right)}. 
		\end{equation}
		
		\item \textbf{Multi-View Convolutional Neural Network Classification on Real-World Data:} The real-world data is based on the ModelNet dataset \cite{su2015multi}, which offers multi-view images of 3D objects.
		Each object example provides $U$ rendered views, and each sensor randomly selects one image of the same class. 
		To reduce transmission overhead, each RGB image is resized from $3 \times 224 \times 224$ to $3 \times 56 \times 56$.
		The MVCNN classifier is based on the VGG11 model \cite{simonyan2015very}, whose backbone is split into a sensor-side feature extractor mainly involving convolutional layers and feature compression, and a server-side classifier mainly involving fully connected layers. 
		Specifically, a $512 \times 1 \times 1$ tensor is generated by the feature extractor and then compressed to a $20 \times 1$ vector.
		In the training, $U$ feature vectors perform arithmetic averaging and the averaged vector is finally processed by the classifier.
		In the testing, local feature vectors from the pre-trained feature extractor experience the AirComp. The aggregated feature vector is finally processed by the pre-trained classifier.
		
	\end{itemize}
	
	\subsubsection{Benchmarks}
	
	To highlight the potential of the proposed design scheme for the edge AI inference in ISCC networks, we include the following benchmarks for comparison purposes. 
	\begin{itemize}
		\item \textbf{Ideal:} The split AI inference model is deployed in an integrated sensing and computation network. The local feature vectors generated by sensor-side feature extractors are directly processed by arithmetic averaging, without communication transmission. 
		Then, this averaged feature vector is fed to the server-side classifier.
		\item \textbf{PAT-MC:} The split AI inference model is deployed in an ISCC network.
		The precoding design scheme is based on communication-centric heuristic methods, where the sensors adopt phase-aligned transmission (PAT) \cite{wang2025ultra}, ${v_{u,k}} = \sqrt {{{\tilde P_u^{{\rm{sensor}}}/ K}}} {{\left| {{\bf{w}}_k^H{{\bf{h}}_{u,k}}} \right|}}/{{{\bf{w}}_k^H{{\bf{h}}_{u,k}}}}, \forall u,k$, and the edge server adopts matched combining (MC), ${{\bf{w}}_k} = \sqrt {P_k^{{\rm{server}}}} {{{{\bf{v}}_k}}}/{{{{\| {{{\bf{v}}_k}} \|}_F}}}$, ${{\bf{v}}_k} = \sum\nolimits_{u = 1}^U {{{\bf{h}}_{u,k}}}, \forall k$.
	\end{itemize}
	
	\subsection{Simulation Results}
	
\begin{figure}[!t]
		\centering
		\subfigure[]{
			\includegraphics[width=0.486\linewidth]{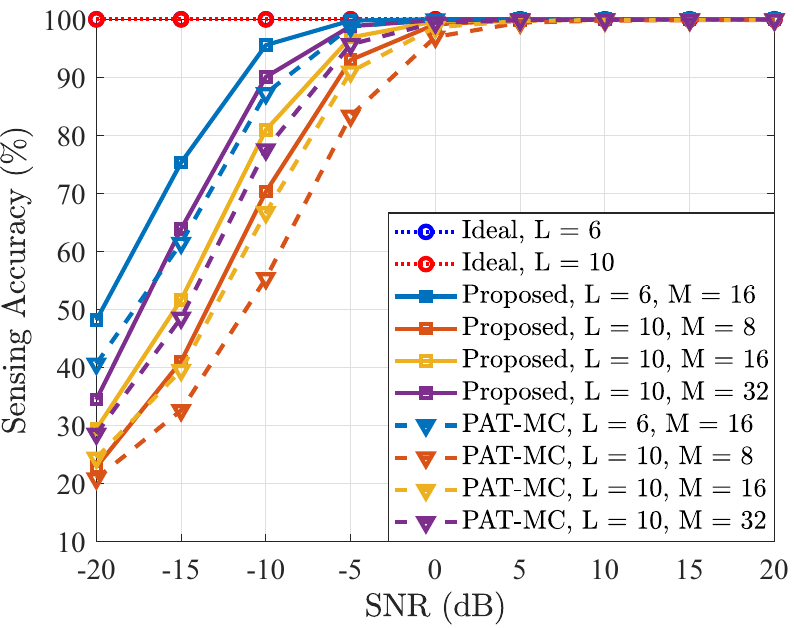} 
			\label{fig:1-1}}
		\hspace{-1.4em}
		\subfigure[]{
			\includegraphics[width=0.486\linewidth]{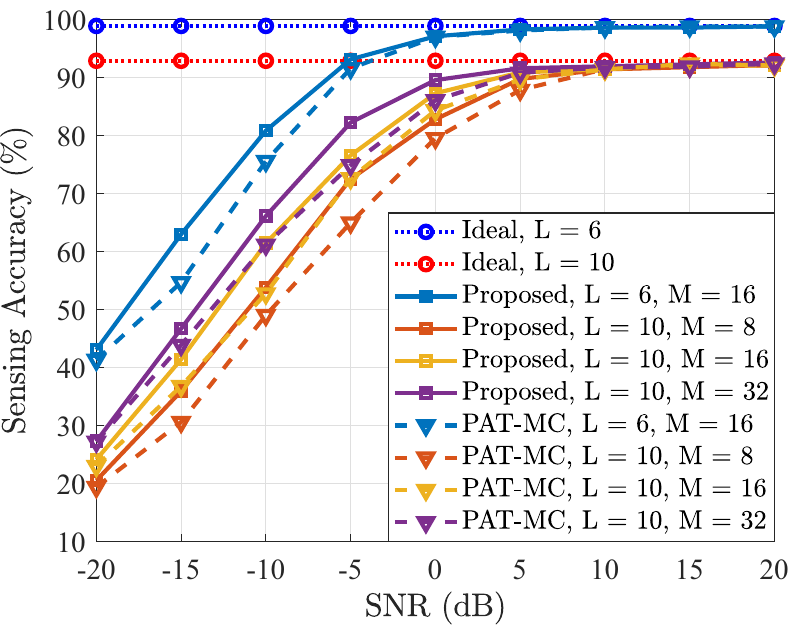} 
			\label{fig:1-2}	}
		\vspace{-1.2em}
		\caption{Sensing accuracy versus ${{\rm{SNR}}}$ across different classification models and datasets. (a) Linear classifier based on synthetic Gaussian mixture dataset; (b) MVCNN classifier based on ModelNet dataset.}
		\vspace{-1.5em}\label{fig:1}
	\end{figure}

		Fig. \ref{fig:1} plots the sensing accuracy versus the analytical equivalent SNR using different classification models and datasets, with $L \in \{6,10\}$ and $M \in \{8,16,32\}$.
		Specifically, Fig. \ref{fig:1-1} compares the performance of the linear classifier on the synthetic Gaussian mixture dataset, while Fig. \ref{fig:1-2} compares the MVCNN classifier on the ModelNet dataset.
		
		We observe that as $\rm SNR$ increases, the sensing accuracy of the proposed design and the PAT-MC increase rapidly at first and then approach the ideal sensing accuracy.
		Notably, in the case $L = 10, M = 16$ in Fig. \ref{fig:1-1}, the proposed design achieves approximately $80\%$ accuracy at $-10$dB $\rm SNR$ while the PAT-MC only reaches about $65\%$, indicating a sensing accuracy improvement of $15\%$.
		Moreover, in the case $L = 10, M = 16$ in Fig. \ref{fig:1-2}, the proposed design achieves approximately $60\%$ accuracy at $-10$dB $\rm SNR$ while the PAT-MC only reaches about $50\%$, indicating a sensing accuracy improvement of $10\%$.
		Furthermore, with $M=16$ fixed, $L=6$ yields higher accuracy than $L=10$, and for a fixed $L=10$, increasing $M$ from $8$ to $32$ causes higher accuracy, with the highest accuracy at $M=32$.
		Therefore, the simulation results validate the effectiveness and feasibility of the proposed design under both synthetic and real-world data with corresponding linear and non-linear classification models, as well as the superiority of the proposed design over the PAT-MC in terms of enhancing sensing accuracy.

	\section{Conclusion}
	
	In this letter, we analyzed sensing accuracy for edge AI inference in an AirComp-empowered ISCC network.
	We exploited the DG to characterize the sensing accuracy and established an explicit functional relationship between the DG and precoding coefficients.
	Building upon this analysis, we formulated a typical sensing accuracy-aware precoding design problem and found efficient solutions that directly maximize the DG.
	Simulation results validates the effectiveness and feasibility of the proposed design in improving sensing accuracy under both synthetic and real-world datasets.
	Despite a moderate increase in computational complexity, the proposed method achieves considerable sensing accuracy gains, especially under low SNR.
	
	\balance
	\newpage
	\bibliographystyle{ieeetr}
	\bibliography{IEEEabrv,ref_v4}
	
	\clearpage
	\setcounter{page}{1}
	
	\section*{Supplementary Material (Appendix)}\label{SM}
	
	\subsection{Proof of Lemma \ref{lemma:1}}\label{app:A}
	\setcounter{equation}{0}
	\renewcommand*{\theequation}{A-\arabic{equation}}
	
	Note that the aggregated feature vector ${\bf{\hat x}}$ is extracted from the estimated symbol \eqref{eq:4}.
	Since the local feature vector for the $l$-th class ${\bf{x}}_u \sim {\cal N}\left( {{{\bm{\mu }}_l},{\bm{\Sigma }}} \right), \forall u$ and communication noise ${\bf{n}}_k \sim {\cal C}{\cal N}\left( {{\bf{0}},\sigma _{\rm{c}}^2{\bf{I}}} \right), \forall k$ follows Gaussian distribution, the PDF of ${\bf{\hat x}}$ for the $l$-th class can be expressed as
	\begin{equation}
		f\left( {{\bf{\hat x}}|l} \right) = {\cal N}( {{{{\bm{\hat \mu }}}_l},{\bf{\hat \Sigma }}}),
	\end{equation}
	where the mean vector ${{{\bm{\hat \mu }}}_l} = {\left[ {{{\hat \mu }_{l,1}}, \ldots,{{\hat \mu }_{l,N}}} \right]^T} \in {{\mathbb R}^N}$, and the covariance matrix ${\bf{\hat \Sigma }} = {\rm{blkdiag}}( {{{{\bf{\hat \Sigma }}}_1}, \ldots ,{{{\bf{\hat \Sigma }}}_K}} ) \in {{\mathbb R}^{N \times N}}$ has a block diagonal structure because different subcarriers are independent of each other, with ${{{\bf{\hat \Sigma }}}_k} = \left[ {\begin{array}{*{20}{c}}
			{{\Sigma _{11,k}}}&{{\Sigma _{12,k}}}\\
			{{\Sigma _{21,k}}}&{{\Sigma _{22,k}}}
	\end{array}} \right] \in {{\mathbb R}^{2 \times 2}}$.
	Based on \eqref{eq:4} and \eqref{eq:5}, the feature element extracted from the received signal at the $k$-th subcarrier is given by 
	\begin{subequations}
		\begin{align}
			\!\!\!{{\hat x}_{2k - 1}} =& \Re \{ {{{\hat s}_k}} \}\!= \!\Re \{ {{\bf{w}}_k^H\textstyle\sum\limits_{u = 1}^U\! {{{\bf{h}}_{u,k}}{v_{u,k}}{s_{u,k}}} \! +\! {\bf{w}}_k^H{{\bf{n}}_k}} \},\\
			{{\hat x}_{2k}} =& \Re \{ {{{\hat s}_k}} \} \!= \! \Im \{ {{\bf{w}}_k^H\textstyle\sum\limits_{u = 1}^U\! {{{\bf{h}}_{u,k}}{v_{u,k}}{s_{u,k}}}  \!+\! {\bf{w}}_k^H{{\bf{n}}_k}}\}.
		\end{align}
	\end{subequations}
	For simplicity, we define ${\omega _{u,k}} = {\bf{w}}_k^H{{\bf{h}}_{u,k}}{v_{u,k}}, \forall u,k$, and then the element of ${{{\bm{\hat \mu }}}_l}$ can be derived as
	\begin{equation}
		\begin{aligned}\label{eq:A3}
			{{\hat \mu }_{l,2k - 1}}\! =&  {\mathbb E}\left[ {{{\hat x}_{2k - 1}}} \right] \\
			=&{\mathbb E}[ {\Re \{ {\textstyle\sum\limits_{u = 1}^U {{\omega _{u,k}}\left( {{x_{u,2k - 1}} + \jmath {x_{u,2k}}} \right)}  + {\bf{w}}_k^H{{\bf{n}}_k}} \}} ]\\
			=&{\mathbb E}[ {\textstyle\sum\limits_{u = 1}^U\!{\Re\!\left\{ {\left( {\Re \left\{ {{\omega _{u,k}}} \right\} \!+\! \jmath \Im \left\{ {{\omega _{u,k}}} \right\}} \right)\left( {{x_{u,2k - 1}} \! + \! \jmath {x_{u,2k}}} \right)} \right\}} } ]\\
			=& \textstyle\sum\limits_{u = 1}^U {\left( {\Re \left\{ {{\omega _{u,k}}} \right\}{\mu _{l,2k - 1}} - \Im \left\{ {{\omega _{u,k}}} \right\}{\mu _{l,2k}}} \right)},
		\end{aligned}\nonumber
	\end{equation}
	and similarly, we can derive ${{\hat \mu }_{l,2k}} ={\mathbb E}\left[ {{{\hat x}_{2k}}} \right]  = \textstyle\sum\nolimits_{u = 1}^U {\left( {\Re \left\{ {{\omega _{u,k}}} \right\}{\mu _{l,2k}} + \Im \left\{ {{\omega _{u,k}}} \right\}{\mu _{l,2k - 1}}} \right)}$.
	Besides, the diagonal element of ${{{\bf{\hat \Sigma }}}_k}$ can be derived as
	\begin{equation}
		\begin{aligned}\label{eq:A5}
			{\Sigma _{11,k}} \!=& {\rm{Cov}}\left[ {{{\hat x}_{2k - 1}},{{\hat x}_{2k - 1}}} \right]
			= {\mathbb E}[ {{{\left( {{{\hat x}_{2k - 1}} -{\mathbb E} \left[ {{{\hat x}_{2k - 1}}} \right]} \right)}^2}} ]\\
			=& {\mathbb E}[ ( \Re \{ {\textstyle\sum\limits_{u = 1}^U {{\omega _{u,k}}( {{x_{u,2k - 1}} + \jmath {x_{u,2k}}} )}  + {\bf{w}}_k^H{{\bf{n}}_k}} \} \\
			&\quad - \textstyle\sum\limits_{u = 1}^U {( {\Re \{ {{\omega _{u,k}}} \}{\mu _{l,2k - 1}} - \Im \left\{ {{\omega _{u,k}}} \right\}{\mu _{l,2k}}} )} )^2 ]\\
			=& {\mathbb E}[ \textstyle\sum\limits_{u = 1}^U ( {{( \Re \left\{ {{\omega _{u,k}}} \right\} )}^2}{( {x_{u,2k - 1}} - {\mu _{l,2k - 1}} )^2} \\
			&\;\;+ {{( {\Im \left\{ {{\omega _{u,k}}} \right\}} )}^2}{{( {{x_{u,2k}} - {\mu _{l,2k}}} )}^2})  ] \!+ \!{\mathbb E}[ (\Re \{ {{\bf{w}}_k^H{{\bf{n}}_k}} \} )^2 ]\\
			=& \!\textstyle\sum\limits_{u = 1}^U \!( {{{\left( {\Re \left\{ {{\omega _{u,k}}} \right\}} \right)}^2}\sigma _{2k - 1}^2 \!+\! {{\left( {\Im \left\{ {{\omega _{u,k}}} \right\}} \right)}^2}\sigma _{2k}^2} ) \! +\! \frac{{\sigma _{\rm{c}}^2}}{2}\left\| {{{\bf{w}}_k}} \right\|_F^2,
		\end{aligned}\nonumber
	\end{equation}
	and similarly, we can derive ${\Sigma _{22,k}} = {\rm{Cov}}\left[ {{{\hat x}_{2k}},{{\hat x}_{2k}}} \right] = {\mathbb E}[ {{{\left( {{{\hat x}_{2k}} - {\mathbb E}[ {{{\hat x}_{2k}}} ]} \right)}^2}} ]
	= \textstyle\sum\nolimits_{u = 1}^U[ {{\left( {\Re \left\{ {{\omega _{u,k}}} \right\}} \right)}^2}\sigma _{2k}^2 + {{\left( {\Im \left\{ {{\omega _{u,k}}} \right\}} \right)}^2}\sigma _{2k - 1}^2] + \ ({{\sigma _{\rm{c}}^2}}/{2})\left\| {{{\bf{w}}_k}} \right\|_F^2$.
	
	Moreover, the off-diagonal element of ${{{\bf{\hat \Sigma }}}_k}$ can be derived as 
	\begin{subequations}
		\begin{align}\label{eq:A7}
			{\Sigma _{12,k}} =& {\Sigma _{21,k}} = {\rm{Cov}}\left[ {{{\hat x}_{2k - 1}},{{\hat x}_{2k}}} \right]\nonumber \\
			=& {\mathbb E}\left[ {\left( {{{\hat x}_{2k - 1}} - {\mathbb E}\left[ {{{\hat x}_{2k - 1}}} \right]} \right)\left( {{{\hat x}_{2k}} - {\mathbb E}\left[ {{{\hat x}_{2k}}} \right]} \right)} \right]\nonumber\\
			=& {\mathbb E}[ ( \Re \{ {\textstyle\sum\limits_{u = 1}^U {{\omega _{u,k}}( {{x_{u,2k - 1}} + \jmath {x_{u,2k}}} )} + {\bf{w}}_k^H{{\bf{n}}_k}} \} \nonumber\\
			&\quad- \textstyle\sum\limits_{u = 1}^U {\left( {\Re \left\{ {{\omega _{u,k}}} \right\}{\mu _{l,2k - 1}} - \Im \left\{ {{\omega _{u,k}}} \right\}{\mu _{l,2k}}} \right)}  )\nonumber\\
			&\quad\cdot  ( \Im \{ \textstyle\sum\limits_{u = 1}^U {{\omega _{u,k}}\left( {{x_{u,2k - 1}} + \jmath {x_{u,2k}}} \right)}  + {\bf{w}}_k^H{{\bf{n}}_k} \} \nonumber\\
			&\qquad-\textstyle\sum\limits_{u = 1}^U {( {\Re \left\{ {{\omega _{u,k}}} \right\}{\mu _{l,2k}} + \Im \left\{ {{\omega _{u,k}}} \right\}{\mu _{l,2k - 1}}} )}  ) ]\nonumber\\
			=& \textstyle\sum\limits_{u = 1}^U \Re \left\{ {{\omega _{u,k}}} \right\}\Im \left\{ {{\omega _{u,k}}} \right\}{\mathbb E}[ {\left( {{x_{u,2k - 1}} - {\mu _{l,2k - 1}}} \right)}^2 ] \nonumber\\
			&-  \textstyle\sum\limits_{u = 1}^U \Im \left\{ {{\omega _{u,k}}} \right\}\Re \left\{ {{\omega _{u,k}}} \right\}{\mathbb E}[ {{{\left( {{x_{u,2k}} - {\mu _{l,2k}}} \right)}^2}} ] \nonumber \\
			=&  \textstyle\sum\limits_{u = 1}^U {\Re \left\{ {{\omega _{u,k}}} \right\}\Im \left\{ {{\omega _{u,k}}} \right\}\left( {\sigma _{2k - 1}^2 - \sigma _{2k}^2} \right)}.\nonumber
		\end{align}
	\end{subequations}
	Based on the above, the feature element pair $\left( {{{\hat x}_{2k - 1}},{{\hat x}_{2k}}} \right),\forall k$ are correlated. 
	For analytical tractability, we apply an approximation $\sigma _{2k - 1}^2 = \sigma _{2k}^2 = \delta _k^2,\forall k$. 
	This leads to ${\Sigma _{12,k}} = {\Sigma _{21,k}} = 0,\forall k$, implying that different dimensions of aggregated feature vector ${\bf{\hat x}}$ of the $l$-th class are statistically independent.
	Then, we can rewrite the corresponding covariance as ${\bf{\hat \Sigma }} = {\rm{diag}}( {{\hat \Sigma }_{1,1}},\dots,{{\hat \Sigma }_{N,N}} ) \in {{\mathbb R}^{N \times N}}$, with ${{\hat \Sigma }_{2k - 1,2k - 1}} = {{\hat \Sigma }_{2k,2k}} = \delta _k^2 \sum\nolimits_{u = 1}^U {{{\left| {{\omega _{u,k}}} \right|}^2}}  + \frac{{\sigma _{\rm{c}}^2}}{2}\left\| {{{\bf{w}}_k}} \right\|_F^2,\forall k$.
	
	\subsection{Efficient Solver for Problem \eqref{eq:P1}}\label{app:B}
	\setcounter{equation}{0}
	\renewcommand*{\theequation}{B-\arabic{equation}}

	To solve the formulated non-convex fractional programming problem, ${\{ {v_{u,k}}\} }$, ${\{ {{\bf{w}}_k}\} }$ can be optimized iteratively as follows.
	
	\subsubsection{Optimization of receive precoder}
	With ${\{ {v_{u,k}}\} }$ fixed, the subproblem of optimizing ${\{ {{\bf{w}}_k}\} }$ is given by
	\begin{equation}\label{eq:18}
		\mathop {\max }\limits_{{{\bf{w}}_k}} \;\frac{{{{\left| {{\bf{w}}_k^H{{\bf{g}}_k}} \right|}^2}}}{{\delta _k^2{\bf{w}}_k^H{{\bf{R}}_k}{{\bf{w}}_k} + \frac{1}{2}\sigma _{\rm{c}}^2\left\| {{{\bf{w}}_k}} \right\|_F^2}},\; {\rm{s.t.}}\; \eqref{eq:P1-c},
	\end{equation}
	where ${{\bf{g}}_k} = \sum\nolimits_{u = 1}^U {{v_{u,k}}{{\bf{h}}_{u,k}}}$ and ${{\bf{R}}_k} = \sum\nolimits_{u = 1}^U {{{\left| {{v_{u,k}}} \right|}^2}{{\bf{h}}_{u,k}}{\bf{h}}_{u,k}^H}$, which is a typical Rayleigh quotient problem with power budget.
	Its optimal close-form solution can be derived as ${\bf{w}}_k^ \star  = \sqrt {P_k^{{\rm{server}}}} {{{{{\bf{\tilde w}}}_k}} \mathord{\left/
			{\vphantom {{{{{\bf{\tilde w}}}_k}} {{{\left\| {{{{\bf{\tilde w}}}_k}} \right\|}_F}}}} \right.
			\kern-\nulldelimiterspace} {{{\left\| {{{{\bf{\tilde w}}}_k}} \right\|}_F}}}$, with ${{{\bf{\tilde w}}}_k} = {( {\delta _k^2{{\bf{R}}_k} + \frac{\sigma _{\rm{c}}^2}{2}{{\bf{I}}_M}} )^{ - 1}}{{\bf{g}}_k},\forall k$.
	
	\subsubsection{Optimization of transmit precoder}
	With ${\{ {{\bf{w}}_k}\} }$ fixed, the subproblem of optimizing ${\{ {v_{u,k}}\} }$ is given by
	\begin{equation}\label{eq:19}
		\mathop {\max }\limits_{\{ {v_{u,k}}\} } \textstyle\sum\limits_{l' = 1}^L {\sum\limits_{l < l'} {\sum\limits_{k = 1}^K {\left( {\frac{{{{\left| {\sum\limits_{u = 1}^U {{\alpha _{u,k}}{v_{u,k}}} } \right|}^2}{\psi _{l,l',k}}}}{{\sum\limits_{u = 1}^U {{\beta _{u,k}}{{\left| {{v_{u,k}}} \right|}^2}}  + {\gamma _k}}}} \right)} } }, \;{\rm s.t.} \;\eqref{eq:P1-b},
	\end{equation}
	where ${\alpha _{u,k}} = {\bf{w}}_k^H{{\bf{h}}_{u,k}}$, ${\beta _{u,k}} = \delta _k^2{\left| {{\alpha _{u,k}}} \right|^2}$ and ${\gamma _k} = \frac{{\sigma _{\rm{c}}^2}}{2}\left\| {{{\bf{w}}_k}} \right\|_F^2$.
	To tackle this fractional programming problem, we introduce an auxiliary variable ${{\chi _k}}$ and rewrite its objective function as $g\left( {\{ {v_{u,k}}\} ,{\{\chi _k}\}} \right) = \sum\nolimits_{l' = 1}^L \sum\nolimits_{l < l'} \sum\nolimits_{k = 1}^K {\psi _{l,l',k}}( 2\Re \{ {\chi _k}\sum\nolimits_{u = 1}^U {{\alpha _{u,k}}{v_{u,k}}}  \} - {{\left| {{\chi _k}} \right|}^2}( \sum\nolimits_{u = 1}^U {\beta _{u,k}}{{\left| {{v_{u,k}}} \right|}^2}  + {\gamma _k} ) )$, where the auxiliary variable is updated by $\chi _k^ \star  = \sum\nolimits_{u = 1}^U {{\alpha _{u,k}}{v_{u,k}}} /( {\sum\nolimits_{u = 1}^U {{\beta _{u,k}}{{\left| {{v_{u,k}}} \right|}^2}}  + {\gamma _k}} )$.
	Accordingly, the problem \eqref{eq:19} can be further transformed to
	\begin{equation}
		\mathop {\max }\limits_{\{ {v_{u,k}}\} } \;\textstyle\sum\limits_{l' = 1}^L {\sum\limits_{l < l'} {\sum\limits_{k = 1}^K {g\left( {\{ {v_{u,k}}\} ,\chi _k^ \star } \right)} } },\;{\rm s.t.}\;\eqref{eq:P1-b},\label{eq:B3}
	\end{equation}
	which is convex that can be easily solved by existing solvers.
	
		\subsection{Complexity and Convergence Analysis}\label{app:C}
		\setcounter{equation}{0}
		\renewcommand*{\theequation}{C-\arabic{equation}}
		
		\subsubsection{Complexity Analysis and Comparison}
		
		We analyze the computational complexity of the proposed alternating optimization-based method and PAT-MC benchmark below.
		\begin{itemize}
			\item \textbf{Proposed Method:} The proposed algorithm is based on alternating optimization. Specifically, the transmit precoder $\{v_{u,k}\}$ and the receive precoder are updated iteratively.
			Updating $\{{\bf w}_k\}$ requires $M \times M$ matrix inversion for $K$ subcarriers, resulting in a complexity of ${\cal O}(KM^2)$.
			Updating $\{ v_{u,k} \}$ involves calculating auxiliary variables and solving the subproblem \eqref{eq:B3}. Since it mainly consists of scalar operations and element-wise power normalization across $U$ sensors and $K$ subcarriers, the complexity is approximately ${\cal O}(UK)$. Thus, the overall complexity of the proposed method is ${\cal O}\left(I\left(KM^2 + UK\right)\right)$, where $I$ is the number of iterations.
			
			\item \textbf{PAT-MC:} At the sensors, the PAT requires computing $v_{u,k} = \sqrt{\tilde P_u^{\rm sensor}/K} \, |{\bf w}_k^H {\bf h}_{u,k}| / ({\bf w}_k^H {\bf h}_{u,k})$ for $U$ sensors and $K$ subcarriers, which involves an $M$-dimensional inner product, yielding a complexity of ${\cal O}(U K M)$. At the edge server, MC computes ${\bf v}_k = \sum_{u=1}^U {\bf h}_{u,k}$ and normalizes it to ${\bf w}_k = \sqrt{P_k^{\rm server}} {\bf v}_k / \|{\bf v}_k\|_F$, resulting in a complexity of ${\cal O}(K U M)$ for all subcarriers. Therefore, the overall complexity of the PAT-MC is ${\cal O}(U K M)$.
		\end{itemize}
		Although the proposed method leads to a higher computational complexity compared to the PAT-MC benchmark, it achieves significant gains in sensing accuracy, particularly under low SNR.
		
		\subsubsection{Convergence Analysis of the Proposed Method}
		
		The proposed alternating optimization-based algorithm aims to maximize the overall DG \eqref{eq:P1-a}, subject to the power budget \eqref{eq:P1-b} and \eqref{eq:P1-c}.
		The analysis of convergence is presented below.
		\subsubsection{Monotonicity}
		For notation simplicity, we define:
		\begin{equation}
			f\left( {{\bf{w}},{\bf{v}}} \right) \buildrel \Delta \over = \frac{2}{{L\left( {L - 1} \right)}}\textstyle\sum\limits_{l' = 1}^L {\sum\limits_{l < l'} {{G_{l,l'}}\left( {\{ {{\bf{w}}_k}\} ,\{ {v_{u,k}}\} } \right)} }.
		\end{equation}
		
		In the $i$-th iteration, the algorithm updates the variables alternatively as follows.
		
		\noindent $\bullet$ Step 1: Fix ${{\bf{v}}^{(i)}}$, update ${{\bf{w}}^{(i + 1)}} = \arg {\max _{\bf{w}}}f\left( {{\bf{w}},{{\bf{v}}^{(i)}}} \right)$. Thus, $f\left( {{{\bf{w}}^{(i + 1)}},{{\bf{v}}^{(i)}}} \right) \ge f\left( {{{\bf{w}}^{(i)}},{{\bf{v}}^{(i)}}} \right)$.
		
		\noindent $\bullet$ Step 2: Fix ${{\bf{w}}^{(i + 1)}}$, update ${{\bf{v}}^{(i + 1)}} = \arg {\max _{\bf{v}}}f\left( {{{\bf{w}}^{(i + 1)}},{\bf{v}}} \right)$. Thus, $f\left( {{{\bf{w}}^{(i + 1)}},{{\bf{v}}^{(i + 1)}}} \right) \ge f\left( {{{\bf{w}}^{(i + 1)}},{{\bf{v}}^{(i)}}} \right)$.
		
		Therefore, $f\left( {{{\bf{w}}^{(i + 1)}},{{\bf{v}}^{(i + 1)}}} \right) \ge f\left( {{{\bf{w}}^{(i)}},{{\bf{v}}^{(i)}}} \right)$, which means that the objective function is monotonically non-decreasing.
		
		\subsubsection{Boundedness} Based on Eq. (14), the pair-wise DG on the $k$-th subcarrier is given by 
		\begin{equation}
			f_{l,l',k}(\mathbf{w}_k, v_{u,k}) = \frac{ \left| \sum_{u=1}^U \mathbf{w}_k^H \mathbf{h}_{u,k} v_{u,k} \right|^2 \left| \Delta d_{l,l',k} \right|^2 }{ \delta_k^2 \sum_{u=1}^U \left| \mathbf{w}_k^H \mathbf{h}_{u,k} v_{u,k} \right|^2 + \frac{\sigma_{c}^2}{2} \|\mathbf{w}_k\|_F^2 }.\label{eq:C2}
		\end{equation}
		Under the transmit power constraint at the sensors, we apply the Cauchy-Schwarz inequality to the numerator:
		\begin{equation}
			\left| \sum_{u=1}^U \mathbf{w}_k^H \mathbf{h}_{u,k} v_{u,k} \right|^2 \le \left( \sum_{u=1}^U \|\mathbf{w}_k^H \mathbf{h}_{u,k}\|^2 \right) \left( \sum_{u=1}^U |v_{u,k}|^2 \right).
		\end{equation}
		Given that the power budget $\sum\nolimits_{u = 1}^U {{{\left| {{v_{u,k}}} \right|}^2}}  \le P_k,\forall u$, with $P_k$ denoting a finite power constraint for all the sensors at the $k$-th subcarrier, the numerator of \eqref{eq:C2} is bounded by:
		\begin{equation}
			\text{Numerator} \le C_1 \cdot \left\| {{{\bf{w}}_k}} \right\|_F^2 \cdot P_k \cdot |\Delta d_{l,l',k}|^2,
		\end{equation}
		where $C_1$ is a constant determined by the maximum eigenvalue of the channel covariance matrices.
		
		The denominator consists of the interference power and the AWGN. Since the interference term $\delta_k^2 \sum_{u=1}^U \left| \mathbf{w}_k^H \mathbf{h}_{u,k} v_{u,k} \right|^2$ is always non-negative, the denominator of \eqref{eq:C2} is strictly lower-bounded by the noise floor:
		\begin{equation}
			\text{Denominator} \ge \frac{\sigma_{c}^2}{2} \|\mathbf{w}_k\|_F^2.
		\end{equation}
		As long as the noise power $\sigma_{c}^2$ is strictly positive ($\sigma_{c}^2 > 0$), the denominator never vanishes for any non-zero $\mathbf{w}_k$.
		
		Therefore, by combining the upper bound of the numerator and the lower bound of the denominator, we obtain:
		\begin{equation}
			f_{l,l',k} \le \frac{C_1 \cdot P_k \cdot |\Delta d_{l,l',k}|^2 \cdot \|\mathbf{w}_k\|_F^2}{\frac{\sigma_{c}^2}{2} \|\mathbf{w}_k\|_F^2} = \frac{2 C_1 P_k|\Delta d_{l,l',k}|^2}{\sigma_{c}^2}.
		\end{equation}
		Since $P_k$, $|\Delta d_{l,l',k}|^2$ and $\sigma_{c}^2>0$ are all finite, each term $f_{l,l',k}$ is globally upper-bounded. 
		Therefore, the overall DG, which is a finite sum of $f_{l,l',k}$, is also upper-bounded, i.e., $\frac{2}{{L\left( {L - 1} \right)}}\sum\nolimits_{l' = 1}^L {\sum\nolimits_{l < l'} {\sum\nolimits_{k = 1}^K {{f_{l,l',k}}} } } \le {C_2} < \infty$, with $C_2$ denoting a finite non-negative value.
		
		Overall, the convergence of the alternating optimization-based algorithm can be guaranteed.

\end{document}